\documentstyle[aps]{revtex}
\newcommand{\ds}{\displaystyle}
\newcommand{\dsf}{\ds\frac}

\newcommand{\beq}{\begin{equation}}
\newcommand{\eeq}{\end{equation}}

\large
\begin{document}
\large

\begin{center}
\Large\bf

STATIONARY THERMOMAGNETIC WAVES IN SUPERCONDUCTORS
\vskip 0.1cm
{\normalsize\bf N.A.\,Taylanov}\\
\vskip 0.1cm
{\large\em Research Institute of Applied Physics,\\
National University of Uzbekistan,\\
Vuzgorodok, Tashkent, 700174, Uzbekistan\\
e-mail: taylanov@iaph.tkt.uz}
\end{center}
\begin{center}
\bf Abstract
\end{center}

\begin{center}
\mbox{\parbox{14cm}{\small  The propagation of a nonlinear stationary
thermomagnetic wave in superconducting media is qualitatively analyzed
in view of the effect of an external magnetic field. The velocity and
the thickness of the wave front are estimated. It is demonstrated that
the inclusion of an external magnetic field affects the thermomagnetic
wave profile only slightly.
}}
\end{center}
\vskip 0.5cm
        The evolution of Joule heat during the dissipative motion of
a magnetic flux gives rise to pronounced nonlinear effects showing up
as various wave processes. These processes define the dynamics of thermal
and electromagnetic perturbations in superconducting media [1-3]. The
stationary propagation of a nonlinear shock thermomagnetic wave in the
resistive state of a superconductor is an example. As is shown [1], the
dispersive, nonlinear, and dissipative processes in superconducting
media result in the formation of stable thermoelectric,$\vec E$, and/or
thermomagnetic, $\vec H$, waves depending on the surface conditions. It was
assumed that the critical current density $\vec j_c$ is independent of the
local value of the external magnetic field $\vec H$ (Bean's model of critical
state [4]). However, in sufficiently strong fields, this dependence
cannot be disregarded (see, e.g., [5]) and its effect on the nonlinear
thermomagnetic wave should be taken into account.
In this paper, we studied the structure of a shock
thermomagnetic wave for an arbitrary dependence of the critical current
density $\vec j_c$ on the external magnetic field $\vec H$. Expressions for the
velocity and the thickness of the wave front were obtained.
The propagation of a thermomagnetic wave in a superconducting medium
along the $x$ axis with a constant velocity $v$ is described by the nonlinear
heat conduction equation (with the self-simulated variables $\xi(x, t) =
x -vt)$ [1]
\beq
-\nu v\dsf{dT}{d\xi}=\dsf{d}{d\xi}\left[\kappa\dsf{dT}{d\xi}\right]+jE,
\eeq
Maxwell equations
\beq
\dsf{dE}{d\xi}=-\dsf{4\pi v}{c^2}j,
\eeq
\beq
E=\dsf{v}{c}H.
\eeq
and the related equation of critical state
\beq
\vec j=j_{á}(T,H)+j_{r}(E),
\eeq
Here, $\nu$ and $\kappa$ are the specific heat capacity and the thermal,
conductivity and $j_r$ is the resistive current density.
In the weak heating approximation $(T-T_0) << (T_c - T_0)$, the resistive
current density $j_r$ in the range of viscous flow ($E > E_f)$;
where $E_f$ is the effective field [5]) is a linear function of
the vortex electric field; i.e., $j_r\approx \sigma_f E$; where $\sigma_f$
is the effective conductivity and
$T_0$ and $T_c$ are the equilibrium and the critical superconductor
temperatures. For weak fields $(E<E_f)$, the dependence
$j_r(E)$ is substantially nonlinear and may be attributed to the thermally
activated motion of the magnetic flux (flux creep [5]).
In the range between two critical magnetic fields $H_{c_1}<<H<<H_{c_2}$
the dependence of the critical current $j_c(T,H)$ on $T$ and $H$ can be
approximated as the product of two functions of single variable:
\beq
j_c=j_c(T,H)=j_c(T)\Phi(H).
\eeq

Here, $j_{c}(T)=j_{c}(T_0)-a(T-T_0)$; $j_{c}(T_0)=j_0$ is the equilibrium
current density, $a$ characterizes the thermal reduction of Abrikosov
vortex pinning on crystal defects, and $\Phi$ will be determined below
for different values of the external field $H$.
The thermal and electrodynamic boundary conditions for Eqs. (1)-(5)
are given by

\beq
\begin{array}{l}
T(\xi\rightarrow+\infty)=T_0, \dsf{dT}{d\xi}(\xi\rightarrow-\infty)=0,\\
\quad\\
E(\xi\rightarrow+\infty)=0,   E(\xi\rightarrow-\infty)=E_e,\\
\end{array}
\eeq
where $E_e$ is the constant external electric field.

      Jointly solving Eqs. (1)-(5) with boundary conditions (6) yields
the following equation for the $E$ wave:
\beq
\dsf{d^2E}{dz^2}+F\left(E,\dsf{dE}{dz}\right)+\dsf{dU}{dE}=0
\eeq
where
\beq
F={\beta(1+\tau)-\dsf{c}{v}\dsf{1}{\Phi}\dsf{d\Phi}{dE}\left[\beta\tau
\dsf{j_r(E)}{\sigma_d}+\dsf{dE}{dz}\right]}\dsf{dE}{dz},
\eeq
\beq
U=\beta^2\tau \int\limits_{0}^{E}{\left[1+\frac{j_0}{j_r}\Phi\right]
\dsf{j_r}{\sigma_d}-\dsf{E^2}{2E_\kappa}\Phi}dE.
\eeq

Here
$z=\dsf{\xi}{L},
\tau=\dsf{4\pi\sigma_f\kappa}{c^2\nu},
\beta=\dsf{vt_\kappa}{L},
t_\kappa=\dsf{\nu L^2}{\kappa},
E_\kappa=\dsf{\kappa}{aL^2},
L=\dsf{cH_e}{4\pi j_0}$ is the magnetic penetration depth,
$\sigma_{d}$  is the differential conductivity, and $H_e$ is the external
magnetic field.

Following [6], let us represent the equation for stationary points in
the phase plane $\left(E,\dsf{dE}{dz}\right)$ as
\beq
\beta^2\tau\dsf{j_r}{\sigma_d}\left[1+\dsf{j_0}{j_r}\Phi\right]-\dsf{E^2}{2E_\kappa}
\Phi=0.
\eeq
In not-too-strong fields, when $H<<H_{c_2}$ the $\Phi\left(\dsf{c}{v}E\right)$
function can be chosen in terms of the Kim-Anderson model [7].
In our situation, Eq. (10) has two equilibrium states: a stable node,
$E_0=0$, and a saddle, $E=E_1$. The phase portrait illustrating these
equilibrium states is plotted in Fig. 1. The separatrix AB, connecting
these two singular points, corresponds to the "overfall"- type solution
with the amplitude $E_e$.
With the boundary conditions $E=E_e$ at $z\rightarrow-\infty$,
this equation can be written as the equation for wave velocity $v_e$:
\beq
1-\Omega_0V_E=X_0(v_E)=\Omega_E v_{E}^{2},
\eeq
where
\beq
\Omega_0=2\tau c\dsf{t_{\kappa}^{2}}{L^2}\dsf{E_\kappa}{H_0}
[1-\dsf{j_0}{\sigma_d E_e}],
\Omega_E=2\tau\dsf{t_{\kappa}^{2}}{L^2}\dsf{E_\kappa}{E_e}.
\eeq

and $H_0$ is a constant parameter.
Common points of the $X_0(v_E)$ function and the quadratic parabola
$\Omega_v {E}^{2}$  yield the velocity $v_E$ we are interested in:

\beq
v_E=\dsf{cE_e}{2H_0}
\left[1-\dsf{j_0}{\sigma_d E_e}\right]
\left[\left(1+\dsf{2L^2}{c^2t_{\kappa}^{2}}
E_\kappa
\dsf{H_{0}^{2}}{\tau E_e}
\left[1-\dsf{j_0}{\sigma_dE_e}\right]^2\right)^{1/2}-1\right].
\eeq

The corresponding plot is depicted in Fig. 2.
Only the positive value $v^{+}$ is physically meaningful.
In the limit  $H_0\rightarrow\infty$ (Bean's model [4]), the wave
velocity is defined by
\beq
v_{E}=\dsf{L}{t_{\kappa}}\left[\dsf{E_{e}}{2\tau E_{\kappa}}\right]^{1/2}.
\eeq

In the reverse case, when $H_0\rightarrow 0$, the velocity is largely
small and is almost independent of the wave amplitude $E_e$:

\beq
v_E\approx \dsf{L^2}{ct_{\kappa}^{2}}
\dsf{H_{0}^{2}}{2\tau E_\kappa E_e}
\left[1-\dsf{j_0}{\sigma_dE_e}\right]^{-1}.
\eeq
For the field $H$ approaching $H_{c_2}$ the following approximation is valid:
\beq
v_E\approx c\dsf{j_0E_e}{j_1H_{c_2}}\left(ln\dsf{E_e}{E_0}\right)^{-1},
\eeq
 In conclusion, our results suggest that taking into account an arbitrary
dependence of $j_c$ ®n $H$ has a minor effect on the wave profile, since
the phase portrait remains nearly the same. The qualitative analysis
presented above may be useful for experimentally studying the penetration
of a magnetic flux into a superconductor placed into an external magnetic
field. Improving the protection of superconducting devices at the design
stage is another possible application of our results.

\begin{center}
 REFERENCE
\end{center}
\begin{enumerate}
\item
I. L. Maksimov, Yu. N. Mastakov, and N. A. Tal1anov, Fiz. Tverd. Tela
(Leningrad) 28,2323 (1986) [Sov. Phys. Solid State 28, 1300 (1986)].
\item
N.A.Taylanov, Superconduc. Science and Technology, 14, 326(2001).
\item
N.A.Taylanov., U.T.Yakhshiev. Technical Phys. Lett.
(Pis'ma v JTF, Russia) 27, 42(2001).
\item
C. P. Bean, Phys. Rev. Lett. 8, 250 (1962).
\item
R.G.Mints and A.L.Rakhmanov, Instability in Superconductors
(Nauka, Moscow, 1984).
\item
V. I. Karpman, Nonlinear Waves in Dispersive Media (Nauka, Moscow, 1973).
\item
P. W. Anderson and Y. B. Kim, Rev. Mod. Phys. 36, 36 (1964).
\end{enumerate}
\end{document}